\documentclass{article}

\usepackage{arxiv}

\usepackage[utf8]{inputenc} 
\usepackage[T1]{fontenc}    
\usepackage{hyperref}       
\usepackage{url}            
\usepackage{booktabs}       
\usepackage{amsfonts}       
\usepackage{nicefrac}       
\usepackage{microtype}      
\usepackage{lipsum}		
\usepackage{graphicx}
\usepackage{natbib}
\usepackage{doi}
\usepackage{physics}
\newcommand{\lword}[1]{\leavevmode\nobreak\hskip0pt plus\linewidth\penalty50\hskip0pt plus-\linewidth\nobreak#1}

\title{Practical insights into controlling exponential concentration in quantum machine learning}

\author{ 
    \href{https://orcid.org/0009-0008-6212-3897}{\includegraphics[scale=0.06]{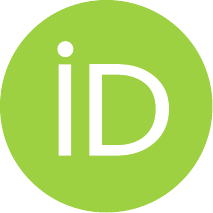}\hspace{1mm}Fernando Mondragon-Sampedro}\\
	CITIC Research Center \\
	Universidade da Coruña \\
	A Coruña, 15071, Spain \\
	\texttt{fernando.mondragon@udc.es} \\
	\And
	\href{https://orcid.org/0000-0001-5790-0577}{\includegraphics[scale=0.06]{orcid.pdf}\hspace{1mm}Diego Alvarez-Estevez } \\
	CITIC Research Center \\
	Universidade da Coruña \\
	A Coruña, 15071, Spain \\
	\texttt{diego.alvareze@udc.es} \\
}

\date{}

\renewcommand{\headeright}{}
\renewcommand{\undertitle}{}
\renewcommand{\shorttitle}{Insights into Exponential Concentration}

\hypersetup{
pdftitle={Practical insights into controlling exponential concentration in quantum machine learning},
pdfsubject={},
pdfauthor={Fernando Mondragon-Sampedro, Diego Alvarez-Estevez},
pdfkeywords={Quantum machine learning \and Exponential concentration \and Quantum kernel methods\and Quantum extreme learning},
}

\begin{document}
\maketitle

\begin{abstract}
Quantum Machine Learning (QML) models are often limited by exponential concentration, where quantum observables become insensitive to input variations as system size increases. In this work, we provide a comparative empirical study of this phenomenon across Fidelity Quantum Kernels (FQK), Projected Quantum Kernels (PQK), and Quantum Extreme Learning (QEL), within a unified observable-based framework based on expectation-value variance.

We investigate how measurement locality, entanglement, and circuit expressivity shape concentration across different architectures. Global fidelity-based kernels exhibit strong concentration driven by Hilbert space dimensionality, whereas local-observable approaches reduce but do not eliminate this effect. In PQK and QEL, concentration is primarily governed by entanglement and quantum dynamical expressivity, with reservoir depth emerging as a key factor in QEL architectures.

We further study bandwidth scaling as a general mechanism for controlling encoding expressivity across all models. Our results indicate that bandwidth scaling effectively reduces concentration in moderate regimes by restricting the effective expressivity of the encoding, though this protective effect diminishes in highly expressive structural settings.
\end{abstract}

\keywords{Quantum machine learning \and Exponential concentration \and Quantum kernel methods\and Quantum extreme learning}

\section{Introduction}
Quantum Machine Learning (QML) studies the use of quantum computational models with the goal of leveraging high-dimensional quantum feature spaces and quantum dynamics for learning and data-processing tasks. In recent years, several paradigms have been developed, including quantum kernel methods \citep{schuld2021supervised}, Quantum Extreme Learning (QEL) architectures \citep{mujal2021opportunities, innocenti2023potential} or Quantum Neural Networks (QNN) \citep{jeswal2019recent}. Many of these approaches are particularly attractive for near-term quantum devices, as they can be implemented using relatively shallow circuits and limited quantum resources. However, despite the growing diversity of QML architectures, scalability remains a central challenge.

As the number of qubits increases, quantum observables can concentrate exponentially around a fixed value, becoming progressively insensitive to variations in the input data. This loss of sensitivity reduces the distinguishability of quantum representations, degrades generalization, and increases the number of measurements required to extract meaningful information. These effects emerge naturally as a consequence of the exponential growth of the high-dimensional induced Hilbert state space \citep{thanasilp2024exponential, xiong2025fundamental}.

This behavior is closely related to the phenomenon of Barren Plateaus (BP) in variational quantum circuits \citep{kairon2025equivalence}, where as the number of qubits grows, the gradient of parameterized quantum circuits vanishes, hindering parameter optimization.

In this work, we provide a comparative study of concentration phenomena across representative quantum machine learning models, namely Fidelity Quantum Kernels (FQK), Projected Quantum Kernels (PQK), and 
QEL architectures. 
We focus on models where exponential concentration directly limits the expressivity and distinguishability of quantum representations via observable expectation values. Consequently, we exclude fully variational architectures, where trainability is predominantly bottlenecked by the gradient-vanishing effects of barren plateaus. Our results highlight how measurement locality, feature map structure, and scaling strategies influence the emergence of exponential concentration, providing practical insights into the design of more scalable QML models.

To enable a unified analysis across different quantum machine learning formulations, we characterize concentration phenomena at the level of observable expectation values. This choice abstracts away from the specific kernel construction for FQK and PQK while preserving the underlying information content, since kernel entries can be expressed in terms of quantum expectation values. This provides a common perspective for comparing different quantum machine learning models under a unified observable-based description.

The remainder of this paper is organized as follows.
Section \ref{secBackground} introduces the concept of exponential concentration in Quantum Machine Learning and discusses its main underlying sources. Section \ref{secMethods} presents the considered QML models and the quantum kernel bandwidth technique. Section \ref{secExperiments} details the experimental methodology, including the datasets, encodings, and evaluation metrics used. Section \ref{secResults} describes the empirical results and analyzes concentration effects across models and configurations. To wrap up, Section \ref{secDiscussion}  discusses the main findings, limitations, and future research directions.

\section{Background}\label{secBackground}
A central motivation of Quantum Machine Learning is the ability of quantum systems to encode information into exponentially large Hilbert spaces, enabling highly expressive feature representations and complex quantum dynamics. However, the scalability of these approaches is hindered by concentration effects that progressively suppress the distinguishability of quantum states and observables as system size increases. This section introduces the theoretical foundations of exponential concentration and the primary mechanisms associated with its emergence in QML models.
\subsection{Exponential concentration}\label{subsecExponentialConcentration}

Quantum observables in high-dimensional systems can exhibit exponential concentration phenomena. Given an observable $O$ acting on an $n$-qubit state $\rho(x)$ encoding a classical input $x \in \mathbb{R}^d$, its expectation value is defined as
\begin{equation}
\langle O \rangle_x = \mathrm{Tr}\!\left[ O \, \rho(x) \right],
\end{equation}
where $\rho(x)$ denotes the density matrix of the quantum state prepared from input $x$.

As the number of qubits increases, the values of $\langle O \rangle_x$ observed across different inputs concentrate around a narrow region centered near a fixed value, reducing their sensitivity to variations in the input data or circuit parameters \citep{thanasilp2024exponential}. As a consequence, different inputs become increasingly difficult to distinguish through measurement outcomes.

While precision limitations can also arise in classical estimation problems, in quantum systems this loss of distinguishability directly impacts measurement complexity. 
Resolving meaningful differences between inputs requires a number of measurements (shots) that scales exponentially with system size, leading to prohibitive resource costs associated with repeated quantum circuit executions.

\subsection{Sources of exponential concentration}\label{subsecSourcesexponentialConcentration}
Several mechanisms have been identified in the literature as potential sources of concentration phenomena in QML models \citep{thanasilp2024exponential}:

\begin{itemize}

    \item \textbf{Circuit expressivity:} Highly expressive quantum circuits can generate state distributions that increasingly resemble those produced by random unitary evolutions. This regime can be formally described using the Haar distribution over unitary operators, which captures the notion of uniformly random quantum evolutions. In high-dimensional Hilbert spaces, Haar-random states exhibit concentration of measure, implying that expectation values of observables cluster around their mean value \citep{mcclean2018barren}. As a consequence, different inputs become increasingly indistinguishable at the level of measurement outcomes, leading to exponential concentration effects.

    \item \textbf{Global measurements:} Global observables act non-locally on the full quantum system, probing correlations across all qubits through a single expectation value. Because these observables attempt to extract information from an exponentially large Hilbert space through a single scalar quantity, expectation values can become increasingly insensitive to structured variations in the input state as system size grows, thereby contributing to concentration phenomena.

    In contrast, local measurements access only reduced subsystems via few-qubit observables, restricting the portion of the Hilbert space that is directly probed by each expectation value.

    \item \textbf{Entanglement:} Strong multi-qubit entanglement in the encoded quantum state can induce concentration effects due to the structure of reduced subsystems. In highly entangled regimes, tracing out qubits leads to reduced density matrices that approach the maximally mixed state, $\rho_A \approx \frac{\mathbb{I}}{2^{n_A}}$, where $n_A$ denotes the number of qubits in subsystem $A$. As entanglement increases, a significant portion of the information about the input state becomes encoded in non-local correlations across the system, causing local observables acting on few-qubit subsystems to become progressively less sensitive to input variations.

    \item \textbf{Hardware noise:} Non-unitary processes, including depolarization and amplitude damping, progressively drive quantum states toward the maximally mixed state. This effectively erases the input-dependent structure encoded within the quantum state. As a result, observable expectation values become increasingly insensitive to the encoded input data, progressively suppressing the information extractable from the model's measurement outcomes.
\end{itemize}

While global measurements and entanglement are often mutually exclusive sources of exponential concentration, since entanglement primarily affects the concentration of locally measured systems, all four factors collectively define the boundaries of trainable QML models.

To study these factors, we analyze a set of representative models that capture and characterize the different sources of exponential concentration. These models are discussed in detail in Section \ref{secMethods}.
 
Noise-induced effects are actively studied in the literature (see, e.g., \cite{agliardi2025mitigating}), but remain outside the scope of this work. Here, we focus on ideal, unitary quantum models to isolate and study concentration effects arising strictly from circuit structure and measurement design.

\section{Methods}\label{secMethods}
In this section, we introduce the Quantum Machine Learning models and mitigation strategies analyzed in this work in the context of exponential concentration phenomena. We first present fidelity-based and projected quantum kernel methods, followed by QEL architectures based on fixed quantum circuit evolution for feature generation.
The relationship between these models and the sources of EC is summarized in Table~\ref{tab:source_solution}.
Finally, we describe quantum kernel bandwidth scaling as a general mechanism for controlling circuit expressivity and concentration effects across different models.

\begin{table*}[h]
\caption{QML models studied and sources of exponential concentration that affect them.}
\label{tab:source_solution}
\centering

\begin{tabular}{lccc}
\toprule
\textbf{Source of EC} &
\textbf{Expressivity} &
\textbf{Global Measurements} &
\textbf{Entanglement} \\
\midrule

Projected Quantum Kernel  & \checkmark &  & \checkmark \\
Quantum Extreme Learning  & \checkmark &  & \checkmark \\
Fidelity Kernel           & \checkmark & \checkmark &  \\

\bottomrule
\end{tabular}

\end{table*}
\subsection{Fidelity Quantum Kernel}\label{subsecFidelityKernel}

Quantum kernel methods extend classical kernel-based learning by embedding classical data into quantum states and defining kernels as inner products in the corresponding induced Hilbert space. Given a quantum feature map $\phi(x)$ for a classical input $x$, the fidelity quantum kernel is defined as
\begin{equation}
K(x_i,x_j) = |\langle \phi(x_i) | \phi(x_j) \rangle|^2,
\end{equation}
where $x_i$ and $x_j$ denote two data points from the dataset.

This kernel function corresponds to the fidelity between quantum states $\ket{\phi(x_i)}$ and $\ket{\phi(x_j)}$, which quantifies the similarity between data points in a high-dimensional quantum Hilbert space \citep{schuld2019quantum, havlivcek2019supervised} The use of fidelity provides a physically meaningful and experimentally accessible measure of similarity, as it can be directly estimated on quantum hardware.
Kernel entries can be computed by preparing the states $\ket{\phi(x_i)}$ and $\ket{\phi(x_j)}$ and estimating their overlap, for example using the Loschmidt Echo method as described in \cite{havlivcek2019supervised}.

The kernel can also be seen as the expectation value of a projector $\Pi_0 = (|0\rangle\langle 0|)^{\otimes n}$ onto the all-zero state:

\[
K(x_i,x_j) = \langle \phi(x_i,x_j) | \Pi_0 | \phi(x_i,x_j) \rangle.
\]
Consequently, the fidelity kernel admits an operational interpretation as the expectation value of a physical measurement: the probability of obtaining the outcome $0^n$ after applying the Loschmidt Echo test and measuring in the computational basis,
\[
K(x_i,x_j) = P(0^n).
\]
where $P(0^n)$ denotes the probability of measuring the all-zero bitstring.
This reformulation allows one to study properties of the kernel, such as concentration phenomena at the observable level. 
Because the observable acts globally over all qubits, it captures non-local correlations across the entire system. This, in turn, bypasses the exponential concentration effects sourced from entanglement, as no subsystems are traced out to yield highly mixed reduced states.

In a machine learning classification context, when combined with classical SVMs, this approach leads to quantum-enhanced SVMs (qSVMs) \citep{schuld2019quantum, havlivcek2019supervised}. More specifically, for a training set $\{(x_i, y_i)\}_{i=1}^N$ with labels $y_i \in \{-1,1\}$, the general SVM decision function is
\begin{equation}
f(x) = \text{sign} \left( \sum_{i=1}^N \alpha_i y_i K(x_i, x) + b \right),
\end{equation}
where the coefficients $\alpha_i$ are obtained by solving the convex quadratic optimization problem
\begin{align}
\max_{\alpha} \quad & \sum_{i=1}^N \alpha_i - \frac{1}{2} \sum_{i,j=1}^N \alpha_i \alpha_j y_i y_j K(x_i,x_j), \\
\text{s.t.} \quad & 0 \leq \alpha_i \leq C, \quad \sum_{i=1}^N \alpha_i y_i = 0.
\end{align}

In this setting, the classical kernel is replaced by the fidelity quantum kernel, enabling the SVM to operate in a quantum-induced feature space  while preserving the convex optimization framework of the classical model \citep{10884820}.

\subsection{Projected Quantum Kernel}\label{subsecProjectedQuantumKernel}

Projected Quantum Kernels were originally introduced to mitigate the concentration effects inherent to global quantum state overlaps, as well as to maximize the geometric difference between the hypothesis spaces of classical and quantum models \citep{huang2021power}. Maximizing this metric is highly desirable because it indicates that the quantum embedding provides a distinct, classically intractable inductive bias.

Instead of comparing full quantum states, PQKs operate on local operators or expectation values of local observables, restricting the measurement space and improving scalability.

A commonly used formulation defines the kernel as

\begin{equation}
k^{\mathrm{PQ}}(x_i, x_j) = \\
\exp\!\left(
-\gamma
\sum_k
\left\|
\rho_k(x_i) - \rho_k(x_j)
\right\|_2^F
\right),
\end{equation}

where $k$ indexes the subsets of one or more qubits chosen to construct the local reduced states, $\rho_k(x)$ denotes the reduced density matrix associated with subsystem $k$, and $\|\cdot\|_2^F$ represents the Frobenius norm.

More generally, projected quantum kernel methods comprise architectures in which classical data are embedded into quantum states and subsequently mapped into a new feature space through the measurement of local observables. 
Standard classical kernel methods, such as SVMs, are then applied to these projected representations.

It is important to note that, although local measurements mitigate the concentration associated with global observables, projected kernels remain susceptible to exponential concentration. As system size increases, the expressivity and entanglement of the encoding can still render local observables completely insensitive to input variations.

\subsection{Quantum Extreme Learning}\label{subsecQEL}

Quantum Extreme Learning models were introduced as an alternative to fully trainable variational quantum circuits, whose optimization can become increasingly difficult as system size and circuit expressivity grow. In particular, highly expressive variational architectures are known to suffer from issues such as vanishing gradients and barren plateaus \citep{mcclean2018barren}, which can severely hinder their trainability. The main idea behind QEL models is to avoid these optimization difficulties by restricting learning to a classical readout layer and keeping the quantum evolution fixed.

In this setting, a quantum circuit acts as a fixed feature map that embeds classical data into a high-dimensional representation space. Given an input sample $x$, the data are encoded into a quantum state, evolved under a fixed circuit, and measured through a set of observables $\{O_i\}$. The resulting feature vector is defined as
\begin{equation}
z_i(x) = \langle O_i \rangle_x,
\end{equation}
where $\langle O_i \rangle_x = \mathrm{Tr}[O_i \rho(x)]$ denotes the expectation value of observable $O_i$ for the encoded input state $\rho(x)$. These features are then processed by a simple trainable classical model, typically linear, for regression or classification tasks.

\begin{figure*}[t] \centering \includegraphics[width=\textwidth]{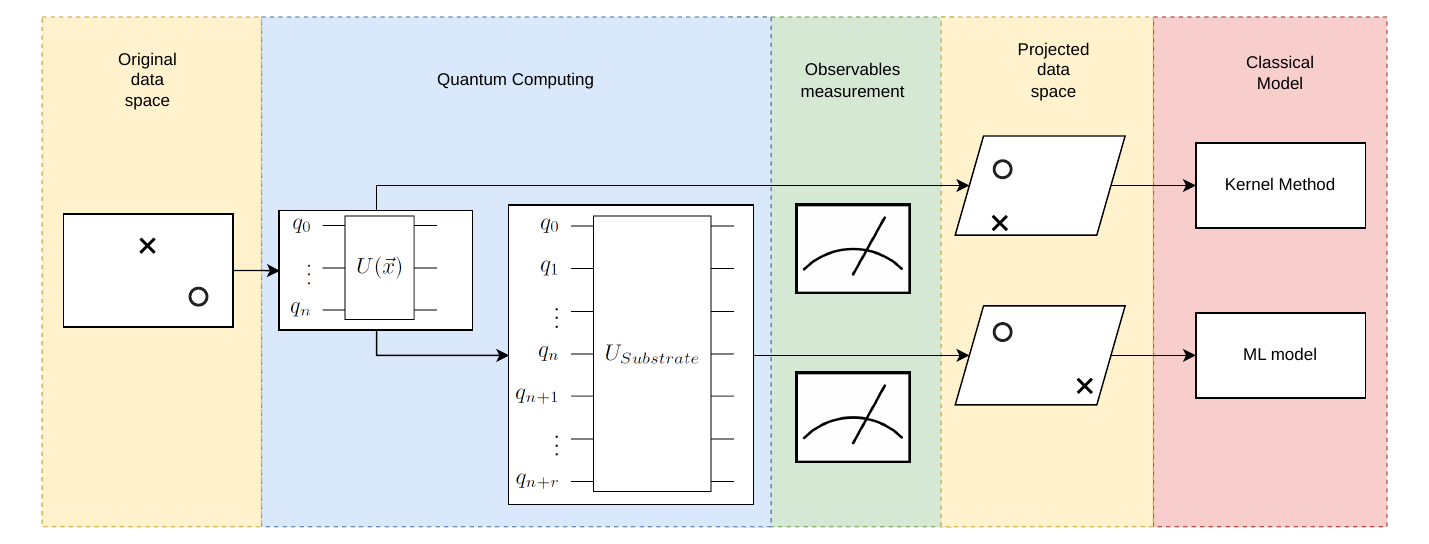} \caption{Comparison between the PQK and QEL pipelines. Yellow blocks represent the different classical data representation spaces, blue blocks correspond to the quantum processing stages, green blocks denote quantum measurement operations, and red blocks represent classical post-processing and learning stages. Both approaches share a common structure consisting of quantum state preparation through the quantum encoding $U(\vec{x})$ acting on $n+1$ qubits, extraction of features via observable measurements, and subsequent classical processing. The main architectural difference is the inclusion of the additional substrate circuit $U_{\mathrm{substrate}}$ in the QEL model, which acts on an enlarged quantum system of $n+r+1$ qubits, where $r$ denotes the number of auxiliary qubits introduced by the substrate}\label{fig:PQKQEL diagram} \end{figure*}

Fig.~\ref{fig:PQKQEL diagram} shows the PQK and QEL model's pipeline, highlighting their shared structure of quantum state preparation, observable-based feature extraction, and classical post-processing. The main difference lies in the quantum feature-processing stage: QEL architectures typically include an additional parameterized unitary circuit called the substrate. This circuit, potentially inducing  higher dimensionality than the encoding circuit, transforms the encoded quantum state prior to measurement. This additional quantum evolution aims to reorganize the embedded information in a way that can be more effectively exploited by the classical learning model.

The QEL formulation is closely related to classical Extreme Learning Machines (ELMs) \citep{huang2004extreme} and reservoir computing frameworks \citep{jaeger2001echo}. Both methods rely on fixed nonlinear transformations followed by a trainable linear readout layer, but they differ fundamentally in feature generation: ELMs use static random hidden layers, whereas reservoir computing extracts features through the continuous evolution of a complex dynamical system. 

The expressive power of QEL architectures is not solely determined by the dimensionality of the induced Hilbert space, but by the interplay between the encoding strategy, the circuit structure, and the choice of measured observables. This decomposition is particularly relevant for scalability analysis, as each component contributes differently to the effective complexity of the induced feature space.

Although QEL architectures circumvent the trainability bottlenecks of barren plateaus, they remain susceptible to the exponential concentration phenomenon because their outputs are derived directly from observable expectation values. As system size and circuit expressivity increase, these observables can become progressively insensitive to input variations, thereby degrading feature distinguishability and limiting scalability. Consequently, balancing expressive feature generation with the risk of exponential concentration is crucial for determining the practical viability of these models.

\subsection{Quantum kernel bandwidth}\label{subsecScalingFactor}

The quantum kernel bandwidth \citep{canatar2022bandwidth} acts as a global scaling factor $\lambda \in (0, 1]$ applied to the input data prior to encoding. This parameter is analogous to the bandwidth parameter ($\sigma$) in classical Radial Basis Function kernels and plays a central role in controlling the inductive bias of FQK models.

Bandwidth scaling can reduce the effective expressivity of quantum feature maps by restricting the region of Hilbert space accessible to the encoding. Consequently, smaller scaling factors may mitigate expressivity-induced exponential concentration by limiting the complexity of the resulting quantum representations.

Although originally motivated in the context of fidelity-based kernel methods, in this work we extend its use to both PQK and QEL models, treating it as a general mechanism for controlling encoding expressivity across architectures. In these settings, scaling acts not only on global state overlaps but also on the statistics of locally measured observables, providing a unified way to probe how concentration depends on input encoding expressivity.

Conversely, restricting the complexity of the induced feature space via excessively small scaling factors can render quantum models classically tractable \citep{florez2025similarity}. Tuning this hyperparameter, therefore, introduces a critical trade-off: one must balance model controllability and expressivity against the need to preserve potential quantum advantage.

\section{Experiments}\label{secExperiments}

In this section, we describe the experimental setup used to evaluate exponential concentration phenomena across Fidelity Quantum Kernels, Projected Quantum Kernels, and Quantum Extreme Learning models. We aim to provide a controlled and consistent comparison of these approaches under varying conditions of encoding expressivity, circuit structure, and system size.

The section is organized as follows. First, we introduce the synthetic dataset used to benchmark the models in a controlled high-dimensional setting. Second, we describe the quantum encodings employed to embed classical data into quantum states, spanning different regimes of expressivity and entanglement. Third, we detail the experimental pipeline and evaluation protocol used to ensure a fair comparison across all models. Finally, we introduce the metrics used to quantify concentration effects and assess model sensitivity to input variations.

\subsection{Dataset}\label{subsecDataset}
To empirically evaluate the practical impact of exponential concentration, we selected a synthetic manifold dataset \citep{goldt2020modeling} consisting of 1,000 data points. The generation process was designed to embed a 2-dimensional signal (the manifold) within a 30-dimensional ambient space. This configuration is specifically chosen to test the models' ability to extract meaningful low-dimensional structures from high-dimensional, noise-heavy inputs. By controlling the generation of the data, we ensure a clear benchmark for how different encodings handle information density and feature reduction under scalable conditions.

\subsection{Encodings}\label{subsecEncodings} 

Quantum encodings are quantum circuits that map classical data into quantum states by embedding input features into parameterized gates. Several encoding strategies have been proposed, differing in how classical information is represented within the quantum state, including amplitude, basis, and angle encodings, among others \citep{schuld2021machine}.

In this work, we focus on angle-based encodings, where classical features are embedded into rotation parameters of single and multi-qubit gates. Choosing the right encoding is a key design choice in quantum machine learning, as it determines how classical information is mapped into the structure of the quantum state, and consequently how it is reflected in the resulting circuit operations and measurements.

In particular, the encoding influences both the locality and entanglement properties of the state, as well as its  expressivity. These aspects affect the behavior of observable expectation values and, in turn, the sensitivity of the model outputs to variations in the input data.

\begin{figure*}[t] \centering \includegraphics[width=\textwidth]{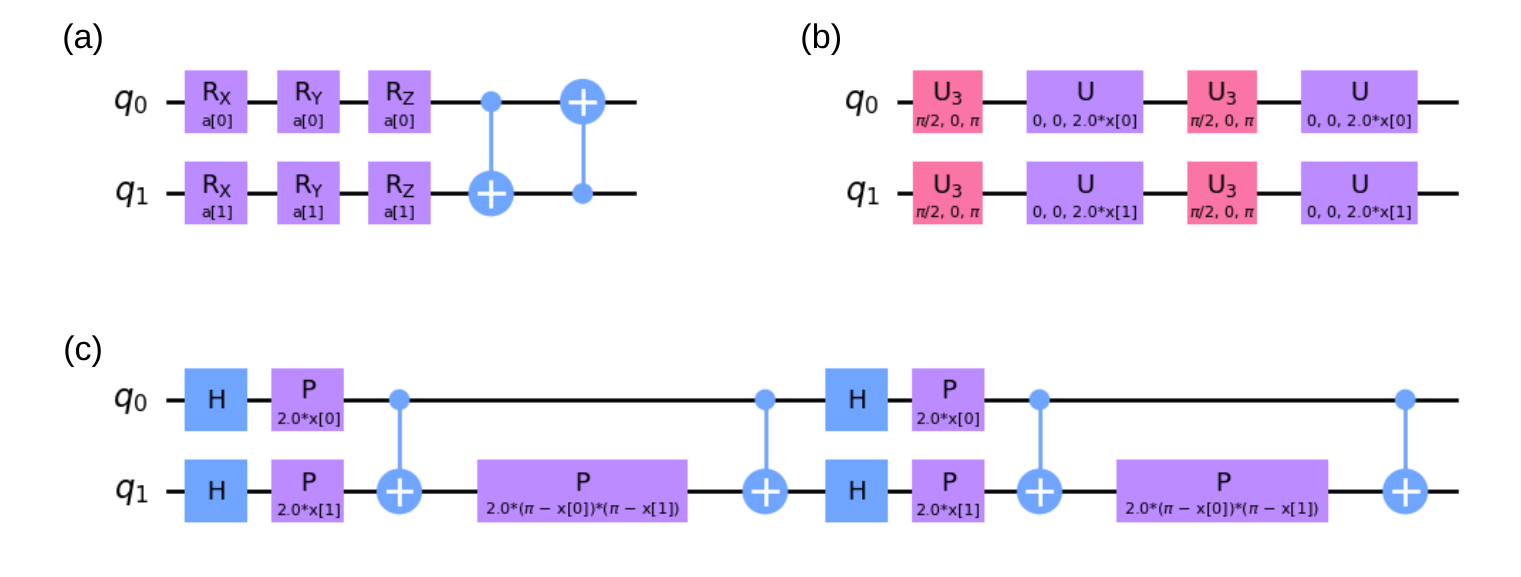} \caption{Encoding circuits used. (a) 3D+CNOT feature map, (b) $Z$ feature map and (c) $ZZ$ feature map.}\label{fig:circuitosUnidos} \end{figure*}

We consider three encodings in our experiments, as illustrated in Fig.~\ref{fig:circuitosUnidos}: 

\begin{itemize}
    \item \textbf{3D+CNOT feature map} \citep{d2024projected}: A multi-axis rotation encoding that combines $R_X$, $R_Y$, and $R_Z$ rotations per qubit with CNOT entangling gates. This construction was originally proposed in the context of PQK methods, aiming to balance simplicity and expressivity while introducing some qubit correlations.

    \item \textbf{Z feature map}: A local angle encoding in which input features are embedded through single-qubit $R_Z$ rotations. As it does not introduce entangling operations, it is used as a no-entanglement baseline to study how exponential concentration emerges under purely local encoding conditions.

    \item \textbf{ZZ feature map}: An extension of the $Z$ feature map that incorporates two-qubit $ZZ$ interaction terms. These entangling operations introduce non-local correlations between encoded features, increasing the expressivity of the quantum state while preserving a structured interaction pattern. This encoding is widely used in the literature and serves as a common benchmark for comparison in quantum machine learning studies \citep{havlivcek2019supervised}.

\end{itemize}

These feature maps were selected to span different levels of locality and entanglement, allowing us to study how encoding-induced state structure affects concentration phenomena in the measured outputs

\subsection{Experimental steps}\label{subsecExperimentalSteps}

We established a unified experimental pipeline to compare the QML models described in Section \ref{secMethods}, namely the FQK, PQK and QEL. The process begins with dimensionality alignment, where the 30-dimensional generated dataset is reduced via Principal Component Analysis (PCA) to match the number of qubits ($n$) used in the encoding for each trial while preserving the most significant variance. The range of qubits used in the experiments varies depending on the model and feature map, as the computational resources needed for each combination differ greatly. The specific covered ranges for each case are specified in Table \ref{Tab:Ranges}. 

\begin{table}[h]
\caption{Ranges of qubits covered by each combination of model and feature map}
\label{Tab:Ranges}
\centering

\begin{tabular*}{\columnwidth}{@{\extracolsep{\fill}}lccc}
\toprule
 & Z & ZZ & 3D+CNOT \\
\midrule
FQK & $2$--$23$ & $2$--$20$ & $2$--$23$ \\
PQK & $2$--$22$ & $2$--$22$ & $2$--$22$ \\
QEL & $2$--$19$ & $2$--$19$ & $2$--$19$ \\
\bottomrule
\end{tabular*}

\end{table}

For the kernel-based methods (FQK and PQK), we conduct 30 independent runs per combination of number of encoding qubits and feature map, utilizing 50 random samples for both training and testing per run to improve statistical robustness. 

The QEL experiments introduce further complexity through the quantum reservoir substrate. This involves varying both the number of auxiliary qubits to expand the total system dimensionality, and the depth of the fixed reservoir unitary to increase the substrates richness. 

We covered a range of auxiliary qubits from 0 to 8, with a maximum total qubit count of 23. As the substrate reservoir, we use random circuits, exploring the range $\{1,2,4,8,16,32,64\}$ as the depth hyperparameter. Due to the higher computational cost associated with simulating these configuration combinations, QEL configurations are evaluated over 3 repetitions instead of 30. 

These protocols are executed across all three feature maps ($Z, ZZ$, and 3D+CNOT) to assess the impact of entanglement and expressivity on the concentration of the outputs.
Models are simulated with a shot budget of 1000 under ideal conditions to isolate the effect of noise from concentration. For the locally measured models (PQK and QEL), we evaluate all single-qubit Pauli observables, namely $\sigma_X$, $\sigma_Y$, and $\sigma_Z$ on each qubit. These observables form a natural local measurement basis and allow us to characterize how concentration develops at the level of individual quantum subsystems.

All experiments are then repeated, now setting the bandwidth scaling factor $\lambda$ = 0.1. This value was chosen based on previous studies showing that scaling factors of this magnitude are sufficient for bandwidth scaling to produce measurable effects \citep{shaydulin2022importance}. 

\subsection{Metrics}\label{subsecExperiments}

Exponential concentration has typically been quantified using the variance of the models' output \citep{thanasilp2024exponential,xiong2025fundamental}. While alternative approaches exist, such as analyzing output probability distributions \citep{aghaei2026pitfalls}, we focus on the average variance of the model outputs across all runs to provide a consistent, robust evaluation criterion across all considered QML architectures.

In this framework, exponential concentration manifests as a reduction in the variability of the measured quantities across different inputs. As observables become increasingly insensitive to input variations, their variance decreases. Consequently, smaller variances indicate stronger concentration effects and a reduced ability of the model outputs to distinguish between inputs.

To enable a uniform comparison, all models are expressed in a common observable-based form. For fidelity-based quantum kernels, kernel entries are interpreted as the expectation value of the global projector onto the all-zero state, $P(0^n)$. For PQK and QEL models, the outputs are directly given by measurements of the single-qubit Pauli observables.

\section{Results}\label{secResults}
This section presents the empirical evaluation of exponential concentration across the three models, both with and without bandwidth scaling. We analyze the evolution of observable variance as a function of system size and encoding strategy.

\subsection{Fidelity Kernel}\label{subsec:ResultsFidelityKernel} 

\begin{figure}[t] \centering \includegraphics[width=\columnwidth]{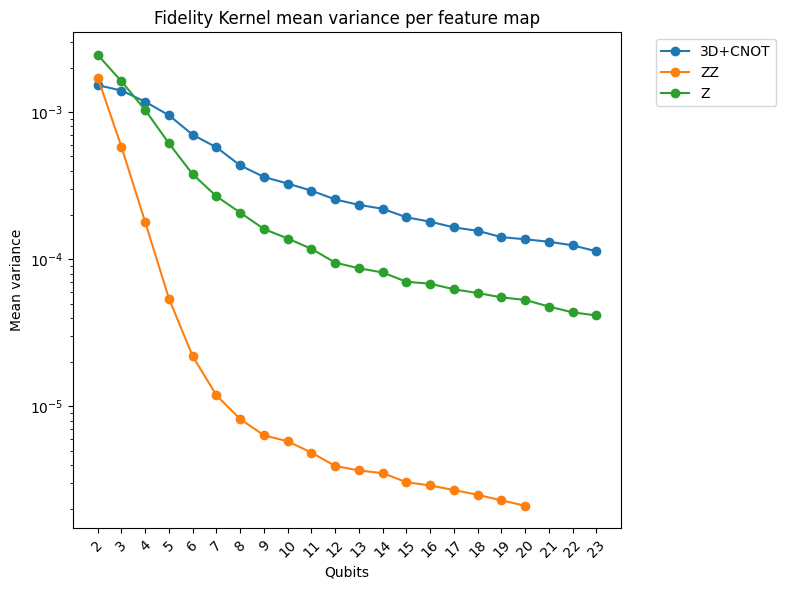} \caption{Mean variance of the fidelity kernel with respect to the number of qubits for the 3 encodings}\label{fig:FK VAR Scaling 1} \end{figure}

As shown in Fig.~\ref{fig:FK VAR Scaling 1}, the mean variance of the projector onto the all-zero state decreases with the number of qubits across all feature maps, indicating the presence of exponential concentration. Among the encodings analyzed, the $ZZ$ feature map exhibits the fastest concentration behavior.

\begin{figure}[t] \centering \includegraphics[width=\columnwidth]{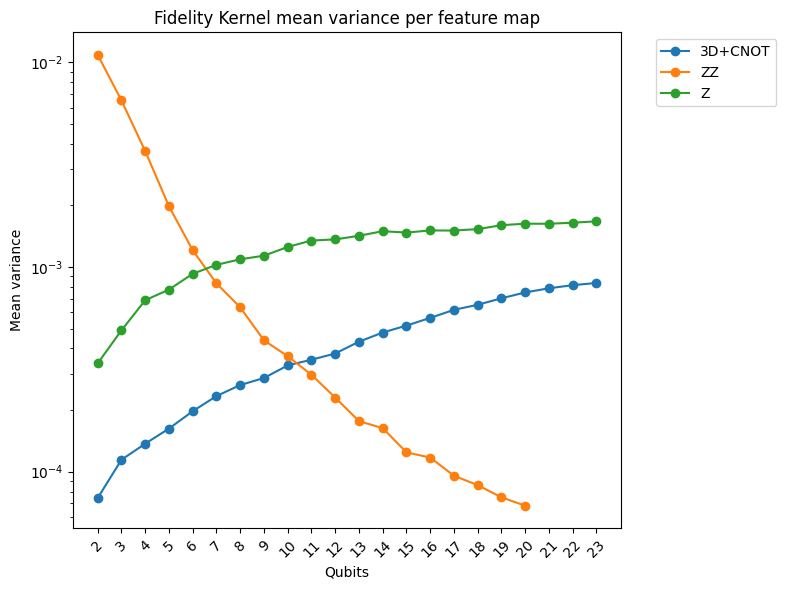} \caption{Mean variance of the fidelity kernel with respect to the number of qubits for the 3 encodings with a bandwidth scaling factor of 0.1}\label{fig:FK VAR Scaling 0.1} \end{figure}

The application of a scaling factor of $\lambda = 0.1$ (Fig.~\ref{fig:FK VAR Scaling 0.1}) mitigates the concentration effects for the $Z$ and 3D+CNOT feature maps, for which no clear exponential decay is observed within the explored regime. In contrast, the $ZZ$ feature map continues to exhibit exponential concentration despite the application of the scaling strategy.

\subsection{Projected Quantum Kernel}\label{subsec:ResultsProjected Quantum Kernel}

\begin{figure*}[t] \centering \includegraphics[width=\textwidth]{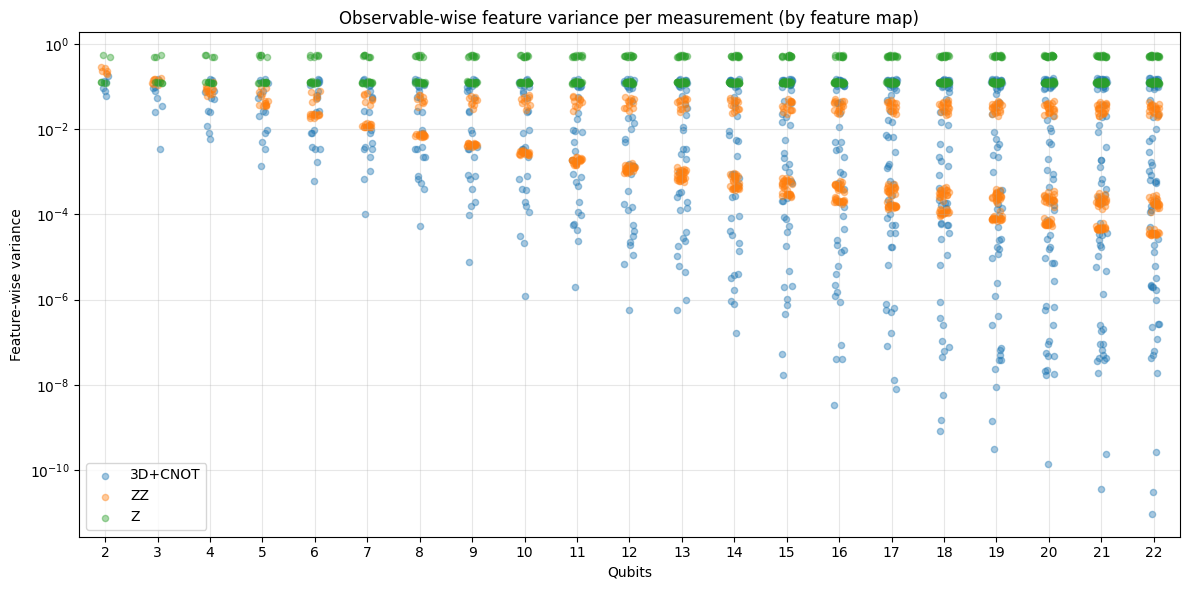} \caption{Mean variance of each single Pauli observables with respect to the number of qubits for the $Z$, $ZZ$, and 3D+CNOT encodings}\label{fig:PQK VAR Scaling 1} \end{figure*}

Fig.~\ref{fig:PQK VAR Scaling 1} shows that the concentration behavior observed in PQK models strongly depends on the encoding. The $Z$ feature map, which does not introduce entanglement, is the only encoding for which the measured observables do not exhibit clear exponential concentration within the explored regime. In contrast, both the $ZZ$ and 3D+CNOT feature maps show a progressive decay in observable variance as the number of qubits increases. Notably, for the 3D-CNOT, different observables exhibit a smooth distribution in their concentration behavior, while for the $ZZ$ encoding, observables cluster into distinct exponentially concentrating regimes.

\begin{figure*}[t] \centering \includegraphics[width=\textwidth]{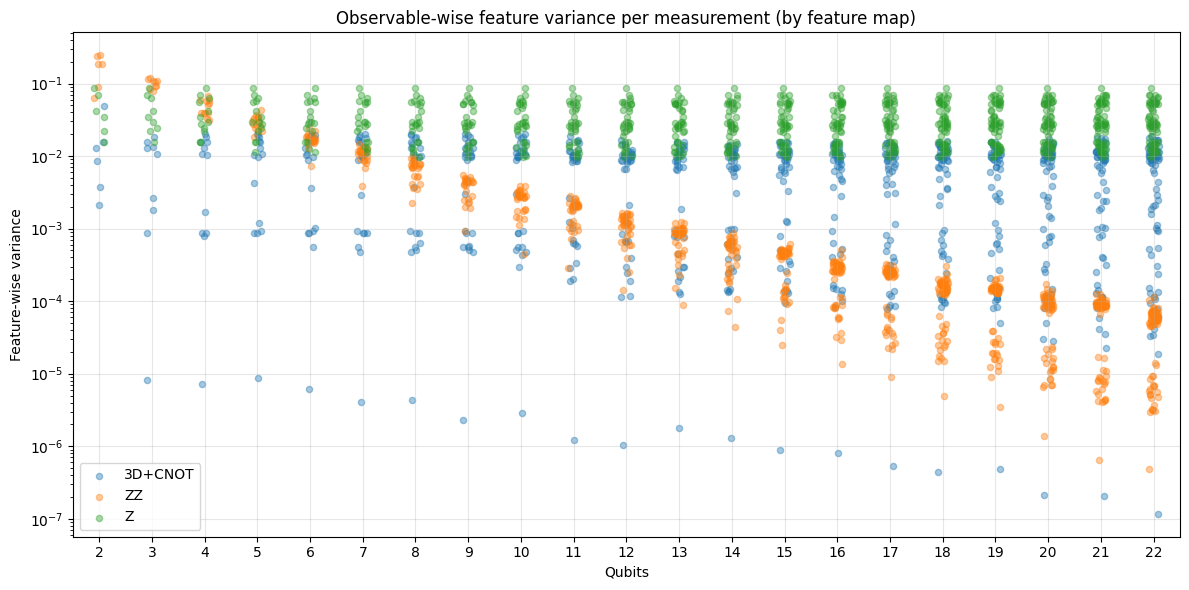} \caption{Mean variance of each single Pauli observables with respect to the number of qubits for the $Z$, $ZZ$, and 3D+CNOT encodings with a bandwidth scaling factor of 0.1}\label{fig:PQK VAR Scaling 0.1} \end{figure*}

As reflected in Fig.~\ref{fig:PQK VAR Scaling 0.1}, the application of the scaling factor yielded qualitatively similar trends. Configurations prone to concentration continued to exhibit variance decay, albeit at a reduced rate. The most prominent difference in the scaled regime is observed in the $ZZ$ feature map, where all observables ultimately succumb to concentration, losing the stability seen in unscaled configurations.

\subsection{Quantum Extreme Learning}\label{subsec:ResultsQuantum Extreme Learning}

Quantum Extreme Learning models exhibit concentration behaviors that depend on both the encoding strategy and the complexity of the reservoir substrate. 

\begin{figure*}[!t] \centering \includegraphics[width=\textwidth]{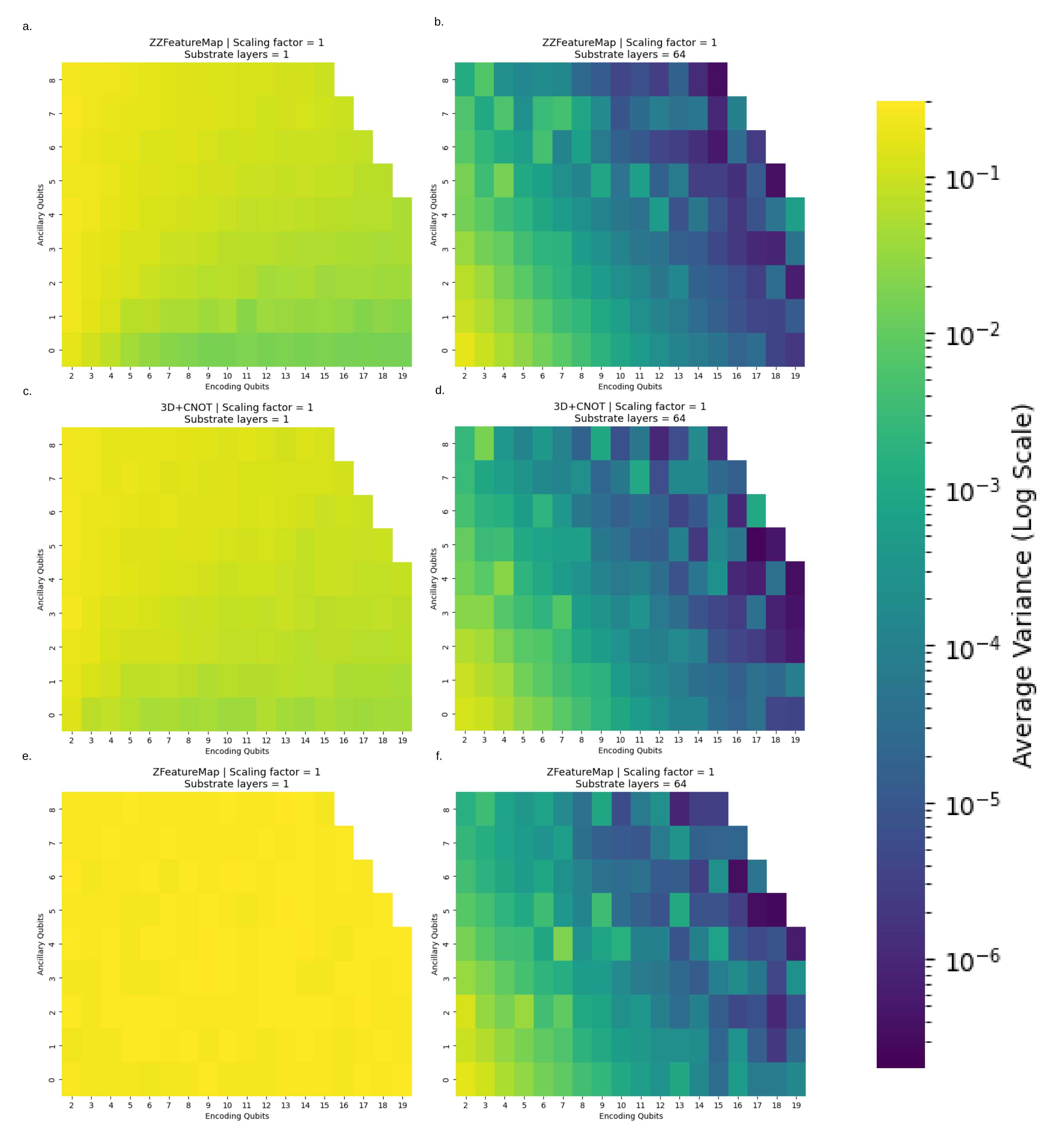} \caption{Mean variance of single-qubit Pauli observables for Quantum Extreme Learning models as a function of the number of encoding qubits in the x-axis and number of ancillary substrate qubits in the y-axis. Each heatmap shows the variance averaged across all observables. Subfigures are grouped by feature map: (a,b), (c,d), and (e,f) correspond to the $Z$, $ZZ$, and 3D+CNOT encodings, respectively. Within each pair, the left panel corresponds to a shallow substrate (1 random layer), while the right panel corresponds to a deep substrate (64 random layers)}\label{fig:QEL avg var scaling 1} \end{figure*}

As illustrated in Fig.~\ref{fig:QEL avg var scaling 1}, for shallow and low-expressivity reservoirs, the observed concentration patterns closely resemble those found in PQK architectures. In this regime, encodings with lower entanglement and reduced expressivity produce observables with larger variance and slower concentration rates, while highly expressive feature maps lead to faster concentration. As reservoir complexity increases,
the inherent concentration dynamics dominate, effectively neutralizing the benefits of the scaling factor.
In particular, increasing reservoir depth produces substantially stronger concentration effects than increasing reservoir width alone. Deep random reservoirs induce rapid observable concentration across all tested encodings.

The addition of ancillary qubits in low-complexity reservoirs does not necessarily accelerate concentration. In several configurations, qubit-expanding architectures preserve observable variance more effectively than fixed-width models with similar substrate depth.

\begin{figure*}[!t] \centering \includegraphics[width=\textwidth]{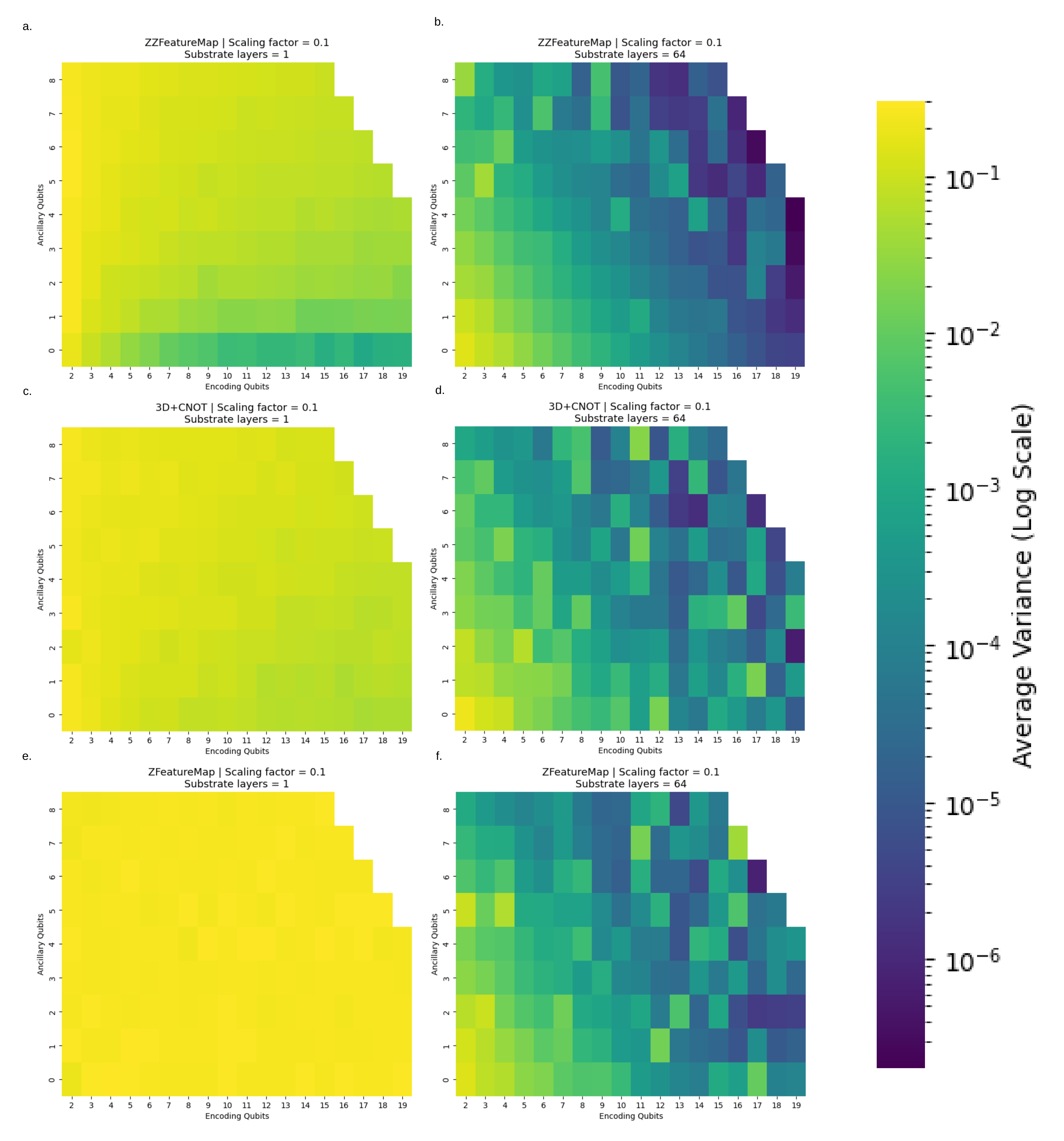} \caption{Mean variance of single-qubit Pauli observables for Quantum Extreme Learning models with a bandwidth scaling factor of $\lambda = 0.1$, as a function of the number of encoding qubits in the x-axis and number of ancillary substrate qubits in the y-axis. Each heatmap shows the variance averaged across all observables. Subfigures are grouped by feature map: (a,b), (c,d), and (e,f) correspond to the $Z$, $ZZ$, and 3D+CNOT encodings, respectively. Within each pair, the left panel corresponds to a shallow substrate (1 random layer), while the right panel corresponds to a deep substrate (64 random layers)}\label{fig:QEL avg var scaling 0.1} \end{figure*}

As illustrated in Fig.~\ref{fig:QEL avg var scaling 0.1}, bandwidth scaling produces noticeably weaker effects in QEL models compared to the FQK and PQK architectures. Although shallow and low-expressivity reservoirs exhibit a slight reduction in concentration when scaling is applied, the observable variance remains strongly affected by the reservoir dynamics. As reservoir expressivity increases, the concentration behavior overcomes the scaling factor. In particular, deep random reservoirs display similar concentration patterns with and without bandwidth scaling.

\section{Discussion}\label{secDiscussion}
The results presented in this work reinforce the idea that exponential concentration constitutes a general scalability limitation across different Quantum Machine Learning paradigms, regardless of whether the model is based on global fidelities, projected observables, or quantum reservoirs. From the observable-based perspective adopted in this study, the main distinction between these architectures lies not in the presence or absence of concentration, but rather in the mechanism through which it emerges.

For FQKs, the concentration of the global projector observable confirms the strong dependence of globally measured architectures on the dimensionality of the Hilbert space. In this setting, highly expressive feature maps accelerate the loss of distinguishability, with the $ZZ$ feature map consistently exhibiting the strongest concentration effects. In contrast, locally measured models such as PQKs and QEL display a more structured behavior, where concentration depends not only on the encoding complexity but also on how entanglement distributes information across observables and quantum subsystems.

The experiments suggest that locality alone is insufficient to guarantee scalable behavior. While local observables alleviate the severe concentration associated with global measurements, rich quantum dynamics and entanglement can still drive observables toward input-independent regimes. This effect becomes particularly evident in QEL architectures, where increasingly deep reservoirs progressively dominate the concentration behavior. Interestingly, reservoir depth appears to play a more significant role than the addition of ancillary qubits, indicating that circuit complexity may be more critical than Hilbert-space dimensionality itself in locally measured models.

One of the clearest trends observed across all architectures is the relationship between encoding expressivity and concentration. Highly expressive encodings consistently produced faster concentration, whereas restricting the effective exploration of the Hilbert space yielded significantly more stable observable variances. In this context, bandwidth scaling emerges as a particularly relevant tool, providing a practical mechanism for controlling feature-map expressivity. Although scaling does not eliminate concentration in highly expressive regimes, the results suggest that limiting the accessible rotational range of encoding circuits can significantly improve observable stability, especially in moderately expressive models. Nevertheless, this mitigation strategy is not without cost. Sufficiently small bandwidth values can make quantum models classically simulable by restricting the complexity of the induced feature space \citep{florez2025similarity}. Consequently, bandwidth scaling introduces a compromise between reducing concentration and preserving the potential for quantum advantage.

Beyond bandwidth scaling, alternative strategies have been proposed in the literature to mitigate concentration phenomena. One approach consists of designing feature maps that exploit prior knowledge about the learning task, for example, by incorporating problem-specific symmetries \citep{henderson2025quantum}. More recently, methods such as Twirlator \citep{uotila2026twirlator} have sought to leverage more general symmetry structures directly at the circuit level, reducing the reliance on task-specific design. These ideas are closely related to the broader framework of geometric quantum machine learning \citep{ragone2022representation}, which studies how geometric and symmetry properties can be exploited to improve quantum representations. In parallel, model-specific techniques such as the Bitflip mitigation method \citep{agliardi2025mitigating} have been shown to reduce noise-induced concentration effects.

The present study has several limitations that should be acknowledged. First, the analysis is conducted using a single synthetic dataset, which may not fully capture the complexity of real-world learning tasks. Second, all experiments are performed under ideal, noiseless simulation conditions, leaving the interplay between exponential concentration and realistic noise sources unaddressed. Moreover, the empirical analysis is restricted to classically tractable system sizes and specific circuit configurations, which may limit the extrapolation of the observed trends to larger-scale regimes. Finally, our conclusions rely on particular choices of observables and concentration metrics, and their connection to downstream learning performance is not explicitly assessed.

In conclusion, our experiments highlight a fundamental trade-off across all analyzed architectures: while increasing expressivity enhances the complexity of the induced quantum feature space, it simultaneously reduces the distinguishability of measured observables. From this perspective, designing scalable QML models requires carefully balanced encoding strategies that avoid both excessively concentrated and overly simplistic representations. More broadly, the observable-level interpretation adopted throughout this work provides a unified framework for analyzing concentration phenomena across architectures that are typically studied in isolation. By offering a comparative perspective on how locality, entanglement, and expressivity interact, this study provides practical insights into mitigating exponential concentration and guiding the development of more scalable quantum machine learning models.

\section{Acknowledgments}\label{secAcknowledgment}

This study has been supported by project \text{RYC2022-038121-I}, funded by \lword{MCIN/AEI/10.13039/501100011033} and European Social Fund Plus (ESF+). CITIC, as a center accredited for excellence within the Galician University System and a member of the CIGUS Network, receives subsidies from the Department of Education, Science, Universities, and Vocational Training of the Xunta de Galicia. Additionally, it is co-financed by the EU through the FEDER Galicia 2021-27 operational program (Ref. ED431G 2023/01).

This study was made possible through the access granted by the Galician Supercomputing Center (CESGA) to its supercomputing infrastructure. The supercomputer FinisTerrae III and its permanent data storage system have been funded by the NextGeneration EU 2021 Recovery, Transformation and Resilience Plan, ICT2021-006904, and also from the Pluriregional Operational Programme of Spain 2014-2020 of the European Regional Development Fund (ERDF), ICTS-2019-02-CESGA-3, and from the State Programme for the Promotion of Scientific and Technical Research of Excellence of the State Plan for Scientific and Technical Research and Innovation 2013-2016 State subprogramme for scientific and technical infrastructures and equipment of ERDF, CESG15-DE-3114.

\bibliographystyle{unsrtnat}
\bibliography{references}  






\end{document}